\documentclass[3p,times,procedia]{elsarticle}
\usepackage{nupha_ecrc}

\volume{00}

\firstpage{1}

\journalname{Nuclear Physics A}

\runauth{J. Churchill et al.}

\jid{nupha}

\jnltitlelogo{Nuclear Physics A}

\usepackage{amssymb}
\usepackage[figuresright]{rotating}

\def\S{{\mathcal S}}

\def\x{\bf x}
\def\p{\bf p}

\def\be{\begin{equation}}
\def\ee{\end{equation}}
\def\bea{\begin{eqnarray}}
\def\eea{\end{eqnarray}}

\usepackage{caption}

\usepackage{titlesec}
\begin{document}

\begin{frontmatter}



%% Instructions from Editor: Please use the following \dochead only in the preprint version (e-print arXiv etc.); 
%% use empty \dochead{} when submitting to Nuclear Physics A!
\dochead{XXVIIIth International Conference on Ultrarelativistic Nucleus-Nucleus Collisions\\ (Quark Matter 2019)}
%% e.g. \dochead{17th International Conference on Dynamical Processes in Excited States of Solids}

\title{Electromagnetic radiation from the pre-equilibrium/pre-hydro stage of the quark-gluon plasma}


%\author[label1,label2,label1,label1]{Jessica Churchill, Li Yan, Sangyong Jeon, Charles Gale}
\author[label1]{Jessica Churchill}
%\affiliation{Department of Physics, McGill University, 3600 University Street, Montreal, QC, Canada H3A 2T8}
\author[label2]{Li Yan}%
%\affiliation{Institute of Modern Physics, Fudan University, Handan Road 220, Yangpu District, Shanghai, 200433, China}
\author[label1]{Sangyong Jeon}
%\affiliation{Department of Physics, McGill University, 3600 University Street, Montreal, QC, Canada H3A 2T8}
\author[label1]{Charles Gale}
%\affiliation{Department of Physics, McGill University, 3600 University Street, Montreal, QC, Canada H3A 2T8}

\address[label1]{Department of Physics, McGill University, 3600 University Street, Montreal, QC, Canada H3A 2T8}
\address[label2]{Institute of Modern Physics, Fudan University, Handan Road 220, Yangpu District, Shanghai, 200433, China}

\begin{abstract}
Photons produced in the pre-equilibrium/pre-hydro stage of the quark-gluon plasma produced in relativistic heavy-ion collisions were computed using parton distribution functions obtained from solutions of the Boltzmann equation. The effect of the initial gluon momentum anisotropy $\xi$ and the dependence on the saturation momentum $Q_s$ was investigated. We see that small $Q_s$ results in a photon yield enhancement, whereas a larger $Q_s$ results in a pre-equilibrium photon suppression, owing to the strict constraint of matching to experimental energy density.
\end{abstract}


\end{frontmatter}

% \linenumbers

\section{Introduction}
\label{sec:intro}

Experiments at the Relativistic Heavy-Ion Collider (RHIC) and the Large Hadron Collider (LHC) have revealed an exotic form of matter: the quark-gluon plasma (QGP) \cite{Jacak:2012dx}. The evolution of QGP can be broadly categorized into pre-equilibrium, hydrodynamics, and hadronization stages. Much research has gone into the hydro and freeze-out stages of this system, but more work is now being devoted to the pre-equilibrium phase, which is the focus of this work. We define pre-equilibrium QGP as a dense system of gluons produced in a time scale of order $t \sim 1/Q_s$, where $Q_s$ is the saturation momentum which characterizes the initial wave functions of the nuclei \cite{MUELLER2000227}. This system evolves as quarks and anti-quarks are created dynamically through gluon fusion. 

In contrast to other probes like QCD jets \cite{Connors:2017ptx}, photons produced in heavy-ion collisions can escape relatively unscathed as the electromagnetic interaction is much weaker than the strong interaction which governs the QGP evolution -- $\alpha/\alpha_s \ll 1$, where $\alpha$ and $\alpha_s$ are the fine structure and strong coupling constant respectively. Therefore, the experimental observable chosen to characterize the initial state of relativistic heavy-ion collisions is photon production \cite{Gale:2018ofa}. To do this, momentum distribution functions of partons are required which appear as solutions of the Boltzmann equation.

%Since the electromagnetic interaction is much weaker than the strong interaction that governs the QGP evolution -- $\alpha/\alpha_s \ll 1$ -- photons produced in heavy-ion collisions can escape without undergoing significant interaction. 
%
%This is in contrast to other probes like QCD jets \cite{Connors:2017ptx} which interact via the strong force. 
%
%Therefore, the experimental observable chosen to characterize the initial state of relativistic heavy-ion collisions is photon production \cite{Gale:2018ofa}. To do this, parton distribution functions are required which appear as solutions of the Boltzmann equation.

\section{\label{sec:boltz} Evolution of pre-equilibrium QGP and Boltzmann Equation}
The parton populations $f_{g/q}$ are calculated by solving the Boltzmann equation 
\begin{eqnarray}
\label{eq:boltz}
\Big(\frac{\partial}{\partial t} + \mathbf{v}\cdot\nabla_x\Big)f_{g/q}(t,\x,\p) 
&=&\mathcal{C}_{g/q}[f_{g}(t,{\x},{\p}),f_{q}(t,\x,\p)],
\end{eqnarray}
where $\mathcal{C}_{g/q}$ are the collision integrals for $2\leftrightarrow2$ scattering processes. Quarks and anti-quarks are described using the same transport equation. Using the framework of the diffusion approximation, which assumes small angle scatterings between constituents, the collision integral is simplified as a Fokker-Planck diffusion term
\begin{eqnarray}
\mathcal{C}_{g/q}[f_g(t,{\x},{\p}),f_q(t,\x,\p)]&=& -\nabla_{\mathbf{p}}\cdot\mathcal{J}_{g/q} + \S_{g/q}
%\mathcal{C}_q[f_g(t,{\x},{\p}),f_q(t,\x,\p)]&=& -\nabla_{\mathbf{p}}\cdot\mathcal{J}_q + \S_q.
\end{eqnarray}
where $\mathcal{J}$ and $\mathcal{S}$ are the effective current and source terms \cite{Blaizot:2014jna}. The early time evolution of the QGP is dominated by a longitudinal expansion, which is described using the Bjorken model. The system is initialized at $t_0Q_s=1$. Pure gluons are described by the gluon distribution function \cite{Romatschke:2003ms} inspired by the colour glass picture 
\begin{eqnarray}\label{eq:gluonIC}
f_g(t_0,p) = f_0 \ \theta\Bigg(1-\frac{\sqrt{p_\perp^2 +  p_z^2\xi^2}}{Q_s}\Bigg)
\end{eqnarray}
with initial anisotropy $\xi$, where $\xi=1.0$ is a perfectly isotropic system. The energy density, initially dominated by gluons, can be manipulated to obtain \cite{Churchill}
\begin{eqnarray}
\frac{dE}{d \eta} = 2 N_c C_F A_T \frac{f_0 Q_s^3}{ \left(2 \pi\right)^2} \frac{1}{\xi} {\cal F} (\xi) 
\label{energy_connection}
\end{eqnarray}
where ${\cal F} (x) = \frac{1}{2} \int_{-1}^{+1} d y \left[ 1 - (1 - x^2) y^2 \right]^{1/2}$ and $A_T$ is the transverse area from Glauber calculation \cite{Loizides:2017ack}. This expression is matched to experimental $dE/d\eta$ to determine the initial gluon population $f_0$. It should be noted that this matching is compatible with fitting to the multiplicities at a late evolution stage \cite{Churchill}. We match to experimentally determined $dE/d\eta$ for three systems: RHIC at 200 GeV, LHC at 2.76 TeV, and LHC at 5.02 TeV. We vary $Q_s$ from 1-2 GeV and look at different $\xi$ values to quantify the effect on the photon yield.

\section{Photon Production in Pre-Equilibrium QGP}
The production rate of photons can be derived starting with the expression 
\begin{eqnarray}
E\frac{d^3R}{d^3p} &=& \int \frac{d^3p_1}{(2\pi)^32E_1}\frac{d^3p_2}{(2\pi)^32E_2}\frac{d^3p_3}{(2\pi)^32E_3} \frac{1}{2(2\pi)^3} |\mathcal{M}|^2(2\pi)^4 \delta^4(P_1+P_2-P_3-P) \nonumber\\
&\hspace{1em}& \times f_1(\mathbf{p}_1)f_2(\mathbf{p}_2)[1 \pm f_3(\mathbf{p}_3)],
\end{eqnarray}
with the degeneracy factor $\mathcal{N}$ absorbed into the amplitude $|\mathcal{M}|^2$. This expression includes both the Compton scattering and quark/anti-quark annihilation channels and is simplified using the small angle approximation. This approximation, which assumes low momentum transfer between scattering particles, gives
\begin{eqnarray}
E\frac{d^3R}{d^3p}= \frac{40}{9\pi^2}\alpha\alpha_s \mathcal{L} f_q(\mathbf{p})\int \frac{d^3p'}{(2\pi)^3}\frac{1}{p'}[f_g(\mathbf{p'}) + f_q(\mathbf{p'})]
\end{eqnarray}
where the Coulomb Logarithm %$\mathcal{L} = log(\Lambda_{UV}/\Lambda_{IR})$
\begin{eqnarray}\label{eq:L}
\mathcal{L} = log\frac{\Lambda_{UV}}{\Lambda_{IR}}
\end{eqnarray}
is a divergent integral related to the strong coupling constant. Realistic simulations take the logarithm $\mathcal{L}$ dynamically by quantifying the UV $q_{max}$ and IR $q_{min}$ cutoffs \cite{PhysRevC.95.054904}. In this calculation, $\mathcal{L}$  is determined numerically by matching to leading order AMY \cite{Arnold:2001ms} rates. Performing a change of variables 
%\begin{eqnarray}\label{eq:pz_tilde}
%\tilde{p_z}=p_{\perp}\sinh(\sinh^{-1}(p_z/p_{\perp})-\eta)
%\end{eqnarray}
to account for a non-zero value of $\eta$, an integration over $\eta$ and $\tau$ gives the final expression for the photon yield. Using in the pre-equilibrium distribution functions computed as described earlier, the pre-equilibrium photon yield is determined. 

%\begin{eqnarray}
%\frac{dN}{dy_p d^2\mathbf{p_{\perp}}} &=& \frac{16 A_T }{3\pi^2}\alpha\alpha_s \mathcal{L} \int \tau d\tau d\eta f_q(p_{\perp}, \tilde{p_z}, \tau) \int \frac{p_{\perp}'dp_{\perp}'dp_z'}{(2\pi)^2}\frac{f_g(p_{\perp}', p_z', \tau) + f_q(p_{\perp}', p_z', \tau)}{\sqrt{p_{\perp}'^2+p_z'^2}}
%\end{eqnarray}

\section{Results}

As mentioned previously, the initial gluon population is computed using Eq. (\ref{energy_connection}) which matches to experimental values \cite{Adler:2004zn, PhysRevLett.109.152303,PhysRevLett.116.222302} of $dE/d\eta$. We look at three cases: RHIC at 200 GeV and LHC at both 2.76 TeV and 5.02 TeV. One goal of this work is to investigate the effect of both $Q_s$ and $\xi$ on the photon yield. Thus, we choose two values of $Q_s$: 1 and 2 GeV, and three values of $\xi$: 1.0, 1.5, 3.7. We use $\xi \geq$ 1.0 which gives $P_L/P_T \leq$ 1.0, where $\xi$ = 1.0 is perfectly isotropic, $\xi$ = 1.5 gives a slight deviation from isotropy, and $\xi$ = 3.7 gives an initial $P_L/P_T$ = 0.1. As $Q_s$ = 1 GeV is a bit low for the LHC energies, we also look at $Q_s$ = 2 GeV. However, note that the value of $f_0 \propto Q_s^{-3}$, which implies that doubling $Q_s$ from 1 to 2 Gev will decrease the value of $f_0$ by a factor of 8 and therefore \textit{lower} the photon yield. 

%\begin{center}
%	\begin{table}[h]
%		\begin{tabular}{c||c||rl||rl}
%			& \textbf{RHIC @ 200 GeV}&  \textbf{LHC}  & \textbf{@ 2.76 TeV} & \textbf{LHC} &  \textbf{@ 5.02 TeV}  \\ 
%			$\xi$ & $Q_s = 1$GeV  & $Q_s = 1$GeV  & $Q_s = 2$GeV & $Q_s = 1$GeV  & $Q_s = 2$GeV  \\ \hline
%			1.0 & $f_0=0.76$ & $f_0=2.66$ & $f_0=0.33$ & $f_0=3.20$ & $f_0=0.40$  \\
%			1.5 & $f_0=1.27$ & $f_0=4.44$ & $f_0=0.56$ & $f_0=5.35$ & $f_0=0.67$  \\
%			3.7 & $f_0=3.47$ & $f_0=12.2$ & $f_0=1.52$ & $f_0=9.58$ & $f_0=1.20$
%		\end{tabular}
%	\end{table}
%\captionof{table}{Values of the initial gluon population $f_0$ computed for various energies, anisotropies ($\eta$), and $Q_s$.}
%\label{table}
%\end{center}

%\begin{figure}[!htbp]
%	\begin{centering}
%		\includegraphics[width=0.85\linewidth]{figures/pQCD+pre-equil_1fmc+hydro_photons_RHIC.pdf}
%		\caption{The pQCD, pre-equilibrium, and hydro, and total photon yield compared to STAR and PHENIX data from RHIC @ $\sqrt{s_{NN}} = 200$ GeV with $Q_s = 1 GeV$.}
%		\label{fig:pQCD+pre-equil+hydro_photons_RHIC}
%	\end{centering}
%\end{figure}

\begin{figure}[!htbp]
	\begin{centering}
		\includegraphics[width=0.9\linewidth]{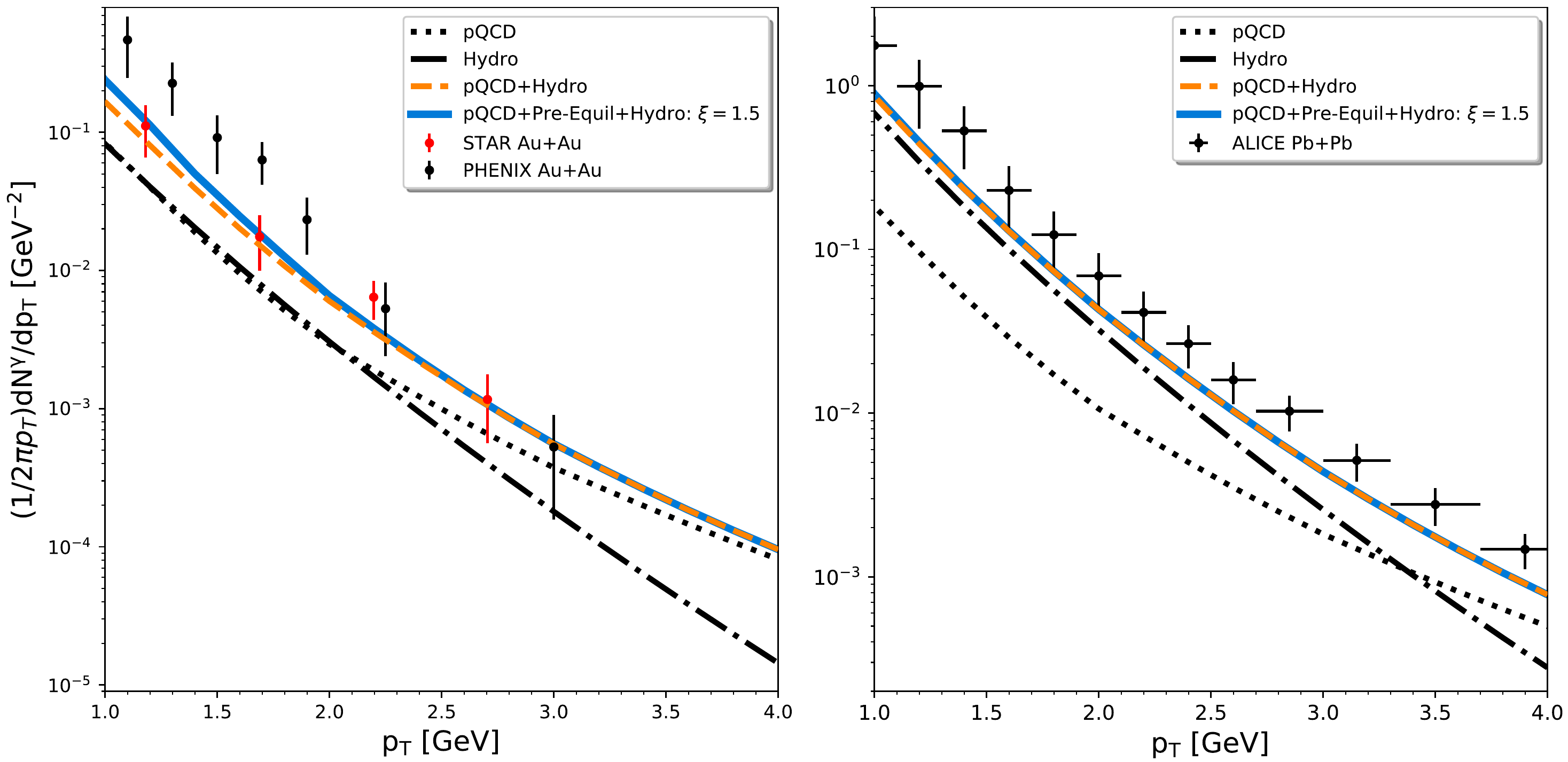}
		\caption{The pQCD, pre-equilibrium, and hydro, and total photon yield compared to STAR and PHENIX data from RHIC @ $\sqrt{s_{NN}} = 200$ GeV with $Q_s = 1$ GeV (left) and ALICE data from LHC at $\sqrt{s_{NN}} = 2.76$ TeV with $Q_s$ = 2 GeV (right).}
		\label{fig:pQCD+pre-equil+hydro_photons_RHIC_and_LHC}
	\end{centering}
\end{figure}

To investigate how the addition of pre-equilibrium photons compares to experimental data from both PHENIX and STAR, hydro and pQCD photons \cite{Paquet} are added. In Fig. \ref{fig:pQCD+pre-equil+hydro_photons_RHIC_and_LHC}, the black dotted lines show the pQCD photons, the black dashed lines show the hydro photons which here have an initialization time of 0.8 fm/c, and the orange dashed lines show the sum of the pQCD and hydro photons. In blue, the pre-equilibrium photons have been added to the sum of the pQCD and hydro photons. On the left side of this figure, we see that this addition increases the total photon yield by about 40\% at $p_T$ = 1 GeV and is in good agreement with STAR data but undershoots PHENIX data.

%\begin{figure}[!htbp]
%	\includegraphics[width=0.85\linewidth]{figures/pQCD+pre-equil_1fmc+hydro_photons_LHC_1_and_2_Qs.pdf}
%	\caption{The pQCD, pre-equilibrium, and hydro, and total photon yield compared to ALICE data from LHC at $\sqrt{s_{NN}} = 2.76$ TeV with $Q_s$ = 1 GeV (left) and $Q_s$ = 2 GeV (right).}
%	\label{fig:pQCD+pre-equil+hydro_photons_LHC}
%\end{figure}

%On the right side of figure \ref{fig:pQCD+pre-equil+hydro_photons_RHIC_and_LHC}, we choose $Q_s$ = 2 GeV and see that the addition of the pre-equilibrium photons increase the total photon yield by ~80\% at $p_T$ = 1 GeV which is in good agreement with experimental data from ALICE. However, since $Q_s$ = 1 GeV is quite low for the LHC, we also look at $Q_s$ = 2 GeV on the right side of figure \ref{fig:pQCD+pre-equil+hydro_photons_RHIC_and_LHC}. In this case, we see a suppression in the photon yield. This is due to the strict constraint imposed by fixing to experimental energy density, which decreases the $f_0$ value as $Q_s$ is increased. Thus, our calculation of photon yield is a conservative estimate.

For the LHC, we see that the addition of the pre-equilibrium photons increase the total photon yield by ~80\% at $p_T$ = 1 GeV which is in good agreement with experimental data from ALICE when taking $Q_s$ = 1 GeV. However, since $Q_s$ = 1 GeV is low for the LHC, we also look at $Q_s$ = 2 GeV shown on the right side of Fig.  \ref{fig:pQCD+pre-equil+hydro_photons_RHIC_and_LHC}. In this case, we see a suppression in the photon yield due to the strict constraint imposed by fixing to experimental energy density, which decreases the $f_0$ value as $Q_s$ is increased.

\begin{figure}[!htbp]
	\begin{centering}
	\includegraphics[width=0.9\linewidth]{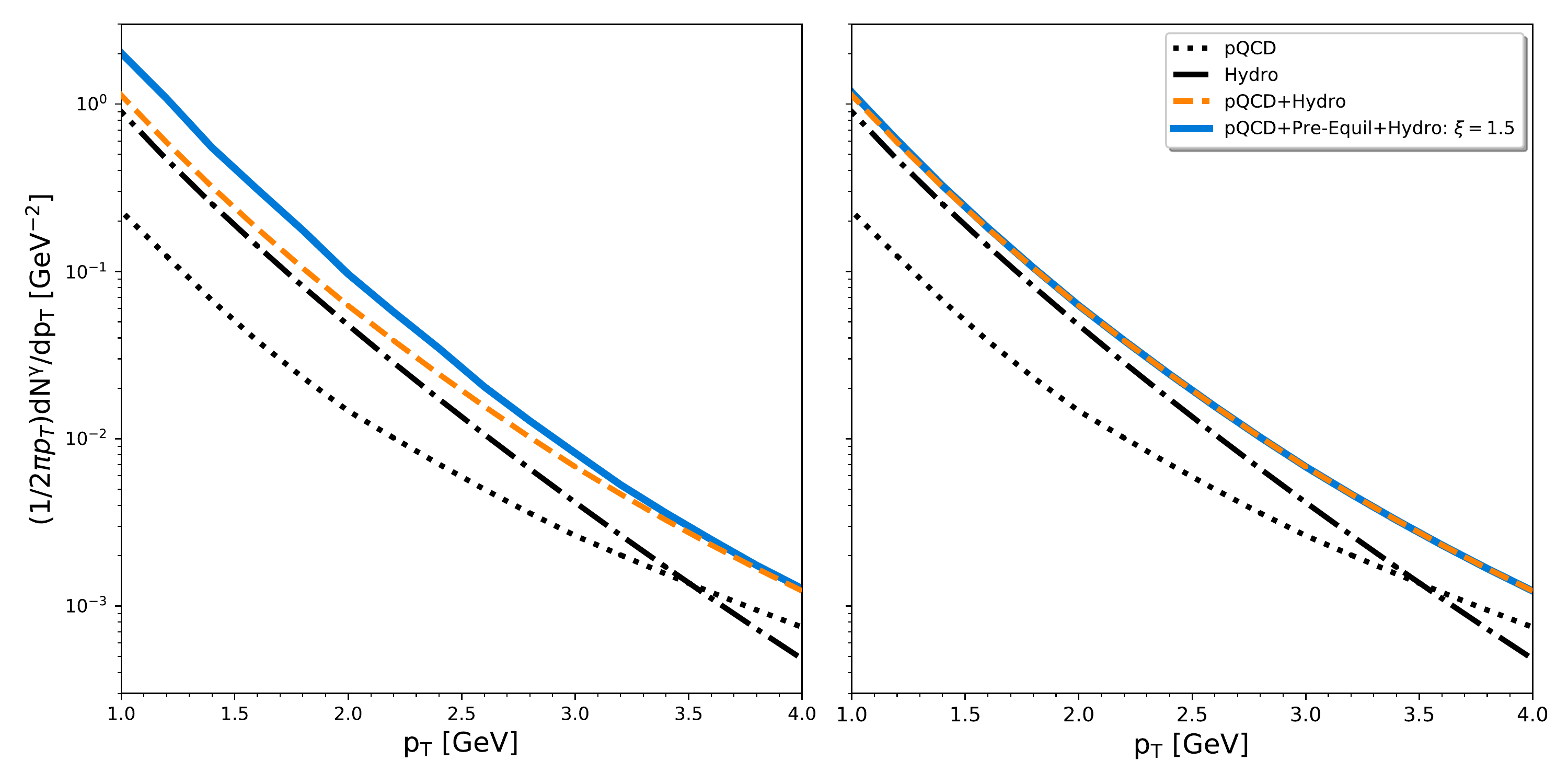}
	\caption{A prediction of the pQCD, pre-equilibrium, and hydro, and total photon yield for LHC at $\sqrt{s_{NN}} = 5.02$ TeV with $Q_s$ = 1 GeV (left) and $Q_s$ = 2 GeV (right).}
	\label{fig:pQCD+pre-equil+hydro_photons_LHC_5.02}
	\end{centering}
\end{figure}

Figure \ref{fig:pQCD+pre-equil+hydro_photons_LHC_5.02} shows a prediction for the photon yield for LHC at $\sqrt{s_{NN}} = 5.02$ TeV. As before, the addition of pre-equilibrium photons enhances the total photon yield by about 80\% when $Q_s$ = 1 GeV. However, when $Q_s$ = 2 GeV, we only see an enhancement of about 4\%. This is again due to fitting our calculation to experimental energy density, which places a strong constraint on the pre-equilibrium photon yield.

\section{Conclusion}

In summary, the photons produced in the pre-equilibrium/pre-hydro stage of QGP have been quantified and the effects of the initial momentum anisotropy as well as the dependence on the saturation momentum $Q_s$ was investigated. When $Q_s$ = 1 GeV, we see that the total photon yield is enhanced by $\sim$40\% at RHIC energies and $\sim$80\% at both LHC energies. However, when $Q_s$ = 2 GeV, we see that the pre-equilibrium photons are suppressed, owing to the strict constraint of matching to experimental energy density, which causes the initial gluon population to decrease by $Q_s^3$. Our results therefore show that the pre-equilibrium photon yield can provide precious information about $Q_s$ and $\xi$. \\

\noindent \textbf{Acknowledgement:} 
This work was supported in part by the Natural Sciences and Engineering Research Council of Canada. C.G. acknowledges useful discussions with L. McLerran, C. Shen, and B. Schenke, J.C. gratefully acknowledges discussions with J.F. Paquet, and all of us are happy to acknowledge discussions with the other members of the Nuclear Theory group at McGill University.



%% References


\bibliographystyle{elsarticle-num}
\bibliography{references}

\begin{thebibliography}{10}
\expandafter\ifx\csname url\endcsname\relax
  \def\url#1{\texttt{#1}}\fi
\expandafter\ifx\csname urlprefix\endcsname\relax\def\urlprefix{URL }\fi
\expandafter\ifx\csname href\endcsname\relax
  \def\href#1#2{#2} \def\path#1{#1}\fi

\bibitem{Jacak:2012dx}
B.~V. Jacak, B.~Muller, {The exploration of hot nuclear matter}, Science 337
  (2012) 310--314.

\bibitem{MUELLER2000227}
A.~Mueller, Toward equilibration in the early stages after a high energy heavy
  ion collision, Nuclear Physics B 572~(1) (2000) 227 -- 240.

\bibitem{Connors:2017ptx}
M.~Connors, C.~Nattrass, R.~Reed, S.~Salur, {Jet measurements in heavy ion
  physics}, Rev. Mod. Phys. 90 (2018) 025005.

\bibitem{Gale:2018ofa}
C.~Gale, {Direct photon production in relativistic heavy-ion collisions - a
  theory update}, in: {12th International Workshop on High-pT Physics in the
  RHIC/LHC Era (HPT 2017) Bergen, Norway, October 2-5, 2017}, 2018.

\bibitem{Blaizot:2014jna}
J.-P. Blaizot, B.~Wu, L.~Yan, Quark production, bose–einstein condensates and
  thermalization of the quark-gluon plasma, Nuclear Physics A 930 (2014) 139 --
  162.

\bibitem{Romatschke:2003ms}
P.~Romatschke, M.~Strickland, {Collective modes of an anisotropic quark gluon
  plasma}, Phys. Rev. D68 (2003) 036004.

\bibitem{Churchill}
J.~Churchill, L.~Yan, S.~Jeon, C.~Gale, The emission of electromagnetic
  radiation from the early stages of relativistic heavy-ion collisions, In
  Preparation.

\bibitem{Loizides:2017ack}
C.~Loizides, J.~Kamin, D.~d'Enterria, {Improved Monte Carlo Glauber predictions
  at present and future nuclear colliders}, Phys. Rev. C97~(5) (2018) 054910.

\bibitem{PhysRevC.95.054904}
J.~Berges, K.~Reygers, N.~Tanji, R.~Venugopalan, Parametric estimate of the
  relative photon yields from the glasma and the quark-gluon plasma in
  heavy-ion collisions, Phys. Rev. C 95 (2017) 054904.

\bibitem{Arnold:2001ms}
P.~B. Arnold, G.~D. Moore, L.~G. Yaffe, {Photon emission from quark gluon
  plasma: Complete leading order results}, JHEP 12 (2001) 009.

\bibitem{Adler:2004zn}
S.~S. Adler, et~al., {Systematic studies of the centrality and s(NN)**(1/2)
  dependence of the d E(T) / d eta and d (N(ch) / d eta in heavy ion collisions
  at mid-rapidity}, Phys. Rev. C71 (2005) 034908.

\bibitem{PhysRevLett.109.152303}
S.~Chatrchyan, et~al., Measurement of the pseudorapidity and centrality
  dependence of the transverse energy density in pb-pb collisions at
  $\sqrt{{s}_{\mathrm{NN}}}=2.76\mathrm{TeV}$, Phys. Rev. Lett. 109 (2012)
  152303.

\bibitem{PhysRevLett.116.222302}
J.~Adam, et~al., Centrality dependence of the charged-particle multiplicity
  density at midrapidity in pb-pb collisions at $\sqrt{{s}_{NN}}=5.02
  \mathrm{TeV}$, Phys. Rev. Lett. 116 (2016) 222302.

\bibitem{Paquet}
J.-F. Paquet, private communication.

\end{thebibliography}


\begin{thebibliography}{}
\expandafter\ifx\csname url\endcsname\relax
  \def\url#1{\texttt{#1}}\fi
\expandafter\ifx\csname urlprefix\endcsname\relax\def\urlprefix{URL }\fi
\expandafter\ifx\csname href\endcsname\relax
  \def\href#1#2{#2} \def\path#1{#1}\fi

\end{thebibliography}






% \end{thebibliography}

\end{document}